\documentclass{elsart}
\usepackage{epsfig,subfigure,latexsym,amsmath}
\begin{document}
\input epsf
\begin{frontmatter}
 
\title{On the Isovector Channels in Relativistic Point Coupling Models within
the Hartree and Hartree-Fock Approximations}
\author[lanl]{D. G. Madland}
\author[lanl]{T. J. B{\"u}rvenich}
\author[fran]{J. A. Maruhn}
\author[erla]{P. -G. Reinhard}
\address[lanl]{Theoretical Division, Los Alamos National Laboratory, Los Alamos,
New Mexico 87545}
\address[fran]{Institute f{\"u}r Theoretische Physik, Universit{\"a}t Frankfurt,
D-60325 Frankfurt, Germany}
\address[erla]{Institute f{\"u}r Theoretische Physik II, Universit{\"a}t
Erlangen-N{\"u}rnberg, D-91058 Erlangen, Germany}
 
\begin{abstract}
We investigate the consequences of Fierz transformations acting upon
the contact interactions for nucleon fields occurring in relativistic
point coupling models in Hartree approximation, which yield the same
models but in Hartree-Fock approximation instead.
Identical nuclear ground state observables
are calculated in the two approximations, but the
magnitudes   of the coupling constants are different.
We find for model studies of four-fermion interactions
occurring in two existing relativistic point coupling phenomenologies
that whereas in Hartree the isovector-scalar
strength $\alpha_{TS}$, corresponding to $\delta$--meson exchange, is unnaturally small,
indicating a possible new symmetry,
  in Hartree-Fock it is instead comparable to the
isovector-vector strength $\alpha_{TV}$ corresponding to   $\rho$--meson
exchange, but the sum of the two isovector coupling constants appears
to be preserved in both approaches.
Furthermore, in Hartree-Fock approximation, both
QCD-scaled isovector coupling constants are {\it natural} (dimensionless
and of order 1) whereas in Hartree approximation only that of
the isovector-vector channel is natural. This indicates that it is not necessary
to search for a new symmetry and, moreover, that the role of the $\delta$--meson
should be reexamined.
This work presents the first comparisons of {\it naturalized} coupling
constants coming from relativistic Hartree and relativistic Hartree-Fock
solutions to the same Lagrangian.
\end{abstract}
 
\begin{keyword}
Relativistic mean-field model \sep Hartree \sep Hartree-Fock \sep Fierz transformation
\PACS 21.10.Dr \sep 21.30.Fe \sep 21.60.Jz \sep 24.85.+p
\end{keyword}
 
\end{frontmatter}
 
\vspace{12pt}
 
\section{Introduction}
 
 Relativistic mean field (RMF) models are quite successful in describing
 ground state properties of finite nuclei and nuclear matter properties.
Such models describe
 the nucleus as a system of Dirac nucleons that interact in a relativistic
 covariant manner via mean meson fields
 \cite{SW86,R89,FST97,LKR97}
 or via mean nucleon fields \cite{Nik92,HMMMNS94,RF97,Bue01}.
The meson fields are of finite range (FR) due
to the meson exchange whereas
the nucleon fields are of zero range (contact interactions or point
couplings PC) together with derivative terms that simulate
the finite range meson exchanges.
A common element to the calculations referenced above is that they
have all been performed in relativistic Hartree approximation.
 
\noindent The RMF-FR studies to date have generally considered three
explicit meson fields.
These are the isoscalar-scalar field due to exchange of
the $\sigma$ meson, the isoscalar-vector field due to
$\omega$ meson exchange, and the isovector-vector field due to
$\rho$ meson exchange.
The isovector-scalar field due to $\delta$
meson exchange has generally not been included because
its contribution to the nuclear force from one-boson exchange is considered
weak \cite{Ma89}, given a relatively large mass of 983 MeV and a relatively
small (but not well determined) coupling constant.
It is, however, included in the RMF-PC studies of Refs.
\cite{Nik92,HMMMNS94,Bue01}, where it is found that its contribution
is very small.
Furthermore, in a study \cite{FML96} of the {\it naturalness} of the set of coupling
coupling constants from Ref. \cite{Nik92}
  it was discovered that the isovector-scalar
coupling constant is unnaturally small. This would presuppose a
symmetry to preserve its small value.
 
\noindent Thus, in relativistic Hartree approximation the isovector-scalar
channel may be neglected, and (perhaps) a symmetry may be identified
to preserve the small value of its coupling constant. However, we have
not found any such symmetry. Therefore, we instead
examine our calculational approach, the relativistic Hartree
approximation, and ask
what happens to the magnitudes of the coupling constants in relativistic Hartree-Fock
approximation where both the direct and exchange terms explicitly
appear?
 
\noindent We believe that the investigation of exchange terms in an
effective field theory for nucleons is meaningful even if one considers the
RMF model as an approximation to the exact density functional in the spirit of the Hohenberg-Kohn theorem
and Kohn-Sham theory \cite{Fur00},
where exchange effects should be absorbed in the
various coupling constants. This is because the Hartree-Fock
theory can be viewed as a Kohn-Sham formalism with exact treatment of exchange
(see Ref. \cite{Dob98}, for example),
which is then a different representation from the Hartree representation.
It also has the correct one-particle limit and
is a self-interaction free theory \cite{Drei90,Dob98}.
This is important for odd systems where the odd particle feels
its own potential if exchange is ignored.
The original idea to perform relativistic Hartree-Fock calculations
by using contact interactions is due to Ref. \cite{MM88}.
 
\noindent In Sec. II we present the simplest possible Lagrangian
containing scalar and vector fields of both isoscalar and isovector
character, and we relate the coupling constants for
  this Lagrangian in relativistic Hartree
approximation to the corresponding coupling constants
  in relativistic Hartree-Fock approximation.
We apply our results in Sec. III to study the four-fermion contact
interactions occurring in two existing realistic point
coupling models determined (phenomenologically) in
relativistic Hartree approximation.
We then address the question of naturalness of the Hartree
and Hartree-Fock coupling constants from these two models
in Sec. IV.
Our conclusions are given in Sec. V.
 
\section{Four-Fermion Relativistic Point  Coupling
with Exchange}
 
For   comparing Hartree and Hartree-Fock representations
we consider two-body contact
interactions (four-fermion point couplings) in the mean field and
no sea approximations, applied to the ground states
of even-even nuclei:
\begin{eqnarray}
{\mathcal{L}} &=&  -\frac{1}{2} \alpha_{  S} ( {\bar{\psi}} {\psi})^2
- \frac{1}{2} \alpha_{  V} ( {\bar{\psi}}\gamma_\mu {\psi}) ( {\bar{\psi}}\gamma^\mu {\psi}) \nonumber \\
&-&\frac{1}{2} \alpha_{  TS} ( {\bar{\psi}}\vec{\tau} {\psi}) \cdot ( {\bar{\psi}}\vec{\tau} {\psi})
- \frac{1}{2} \alpha_{  TV} ( {\bar{\psi}}\gamma_\mu\vec{\tau} {\psi})\cdot ( {\bar{\psi}}\gamma^\mu\vec{\tau} {\psi})
\label{L1}
\end{eqnarray}
 
\noindent where $\psi$ is the nucleon field and $\vec{\tau}$ is the isospin
matrix. We wish to compare the {\em same model ansatz} in two
different many-body approximations. Thus, our model space does not explicitly include pions because
the pion field vanishes
in the Hartree approximation, but contributes via its exchange terms in the Hartree-Fock approximation,
which would then
yield two different models.
Implicitly the effects of the pion are nevertheless
included because our coupling constants are determined by measured observables.
Accordingly, we regard this work as a model study and do not
construct a complete
Hartree-Fock model.
 
\noindent Taking the normal ordered expectation value of $\mathcal{L}$ in a Slater determinant $|\Phi\rangle$
leads to the well-known direct and exchange terms that are to be solved in
Hartree-Fock approximation, namely,
 
\begin{eqnarray}
\langle\Phi| : {\mathcal{L_{\rm HF}}} : |\Phi\rangle &=& -\frac{1}{2} \alpha_{S} {\rho_{S}}^2 - \frac{1}{2}
 \alpha_{V} {\rho_{V}}^2
-\frac{1}{2} \alpha_{TS} {\rho_{TS}}^2 - \frac{1}{2} \alpha_{TV} {\rho_{TV}}^2
 \nonumber \\
&+&\frac{1}{2} \alpha_{S} {\rho_{S_{ex}}}^2 + \frac{1}{2} \alpha_{V} {\rho_{V_{ex}}}^2
+ \frac{1}{2} \alpha_{TS} {\rho_{TS_{ex}}}^2 + \frac{1}{2} \alpha_{TV}
{\rho_{TV_{ex}}}^2
\label{L2}
\end{eqnarray}
 
\noindent Here, we refer to the $\{\alpha\}$ as coupling constants,
$\rho_{S}$ and $\rho_{V}$ denote the isoscalar scalar density and
the time component of the isoscalar vector density, respectively, and
$\rho_{TS}$ and $\rho_{TV}$ denote the corresponding isovector densities.
The squares of the isoscalar scalar and vector exchange densities, ${\rho_{S_{ex}}}$
and ${\rho_{V_{ex}}}$, and those of the isovector scalar and vector exchange densities,
${\rho_{TS_{ex}}}$ and ${\rho_{TV_{ex}}}$, are given by
[$a$ and $b$ are nucleon states]:
 
\begin{eqnarray}
{\rho_{S_{ex}}}^2 &=&  \sum_{a,b} (\bar{\psi}_a\psi_b)(\bar{\psi}_b\psi_a) \\
{\rho_{V_{ex}}}^2 &=&  \sum_{a,b} (\bar{\psi}_a\gamma_\mu\psi_b)(\bar{\psi}_b\gamma^\mu\psi_a) \\
{\rho_{TS_{ex}}}^2 &=& \sum_{a,b} (\bar{\psi}_a\vec{\tau}\psi_b)\cdot(\bar{\psi}_b\vec{\tau}\psi_a) \\
{\rho_{TV_{ex}}}^2 &=& \sum_{a,b} (\bar{\psi}_a\gamma_\mu\vec{\tau}\psi_b)\cdot(\bar{\psi}_b\gamma^\mu
\vec{\tau}\psi_a)
\end{eqnarray}
 
\noindent We then apply Fierz transformations \cite{Fi37,Mar01} in Dirac-iso space
to the four exchange densities.
This transformation expresses a product of nondiagonal matrix elements
of Dirac $\Gamma$-matrices as an expansion into products of diagonal matrix
elements, such as
 
\begin{equation}
(\bar{\psi}_a\Gamma_i\psi_b)(\bar{\psi}_b\Gamma_j\psi_a) = \sum_{k,l=1}^{16}
c_{kl}(\bar{\psi}_a\Gamma_k\psi_a)(\bar{\psi}_b\Gamma_l\psi_b)
\end{equation}
 
\noindent where $\Gamma_i$ stands for one of the sixteen Dirac matrices
$\{{\bf 1}, \gamma_{\mu}, \gamma_5, \gamma_5\gamma_{\mu}, \sigma_{\mu\nu}\}$ constituting
a linearly independent basis in the space of complex 4$\times$4 matrices, which may be
coupled or uncoupled to isospin matrices $\vec{\tau}$.
Applying Eq. (7) to Eqs. (3-6), one immediately sees that all terms containing
$\gamma_5$ and $\gamma_5\gamma_{\mu}$ vanish because $\psi_a$ and $\psi_b$ are
nucleon fields with good parity. We now make the (reasonable) approximation
that the tensor ($\sigma_{\mu\nu}$) and iso-tensor ($\vec{\tau}\sigma_{\mu\nu}$)
contributions are quite small and can be neglected \cite{RF97}.
Reordering the resulting
terms leads to a Lagrangian that is formally identical to
Eq. (\ref{L2}) without the four exchange terms, but with newly defined
coupling constants instead:
 
\begin{equation}
{\mathcal{L}}_{\widetilde{\rm HF}}  = -\frac{1}{2} \tilde{\alpha}_{S} {\rho_{S}}^2 - \frac{1}{2}
 \tilde{\alpha}_{V} {\rho_{V}}^2
- \frac{1}{2} \tilde{\alpha}_{TS} {\rho_{TS}}^2 - \frac{1}{2} \tilde{\alpha}_{TV}
{\rho_{TV}}^2
\label{L3}
\end{equation}
 
\noindent where the newly defined coupling constants are given by
 
\begin{eqnarray}
\tilde{\alpha}_{  S} &\equiv& \;\;\; \frac{7}{8} \alpha_{  S} - \frac{1}{2}
 \alpha_{  V} - \frac{3}{8} \alpha_{  TS} - \frac{3}{2} \alpha_{  TV}
 \nonumber \\
\tilde{\alpha}_{  V} &\equiv& - \frac{1}{8} \alpha_{  S} + \frac{5}{4}
 \alpha_{V} - \frac{3}{8} \alpha_{  TS} + \frac{3}{4} \alpha_{
 TV} \nonumber\\
\tilde{\alpha}_{TS} &\equiv& - \frac{1}{8} \alpha_{S} - \frac{1}{2}
 \alpha_{V} + \frac{9}{8} \alpha_{TS} + \frac{1}{2} \alpha_{
 TV} \nonumber\\
\tilde{\alpha}_{  TV} &\equiv& - \frac{1}{8} \alpha_{S} + \frac{1}{4}
 \alpha_{V} + \frac{1}{8} \alpha_{TS} + \frac{3}{4} \alpha_{TV}
\label{new-alpha}
\end{eqnarray}
 
\noindent This result already shows that in the Hartree-Fock approach, due to the
 exchange effect, all original terms contribute to all channels of the effective
 interaction. {\it And the formal structure of Eq. (\ref{L3}) is identical
to the Hartree approximation for the same
 model with redefined coupling constants.} However, this Lagrangian,
 when considering all terms arising from the Fierz transformations, is a
 self-interaction free theory.
  The inverse solution of Eq. (\ref{new-alpha}) is
\begin{eqnarray}
\alpha_{  S} &=& \frac{34}{21} \,\tilde{\alpha}_{S} + \frac{4}{21}
 \,\tilde{\alpha}_{V} +\frac{6}{21} \,\tilde{\alpha}_{TS} + \frac{60}{21}
 \,\tilde{\alpha}_{TV} \nonumber \\
\alpha_{  V} & = & \frac{1}{21} \,\tilde{\alpha}_{  S} + \frac{31}{21}
 \,\tilde{\alpha}_{V} + \frac{15}{21} \,\tilde{\alpha}_{  TS}
- \frac{39}{21} \,\tilde{\alpha}_{TV} \nonumber \\
\alpha_{  TS} & = & \frac{2}{21} \,\tilde{\alpha}_{S} + \frac{20}{21}
 \,\tilde{\alpha}_{V}
+\frac{30}{21} \,\tilde{\alpha}_{TS} - \frac{36}{21} \,\tilde{\alpha}_{TV} \nonumber
 \\
\alpha_{  TV} & = & \frac{5}{21}\, \tilde{\alpha}_{S} - \frac{13}{21}
 \,\tilde{\alpha}_{V} - \frac{9}{21}\, \tilde{\alpha}_{TS} +\frac{57}{21}\,
 \tilde{\alpha}_{TV}
\label{inverse}
\end{eqnarray}
 
\noindent Given the above results, if one determines the coupling
constants of a relativistic point coupling Lagrangian in Hartree
approximation,
  then the set of coupling constants \{$\tilde{\alpha}$\}
in Eqs. (\ref{L3}) and (\ref{new-alpha}) has been determined.
Use of these coupling constants in Eq. (\ref{inverse}) then
yields the original Hartree-Fock coupling constants
\{$\alpha$\}.
The two sets of coupling constants yield identical predictions of the
nuclear ground state observables, but their magnitudes and physical
interpretation are different because the former set {\it implicitly} accounts
for exchange processes whereas the latter set {\it explicitly} accounts for
exchange processes.
 
\section{Exchange Effects in   Relativistic
Point   Coupling Models
Determined from Measured Observables in Hartree Approximation}
 
We examine coupling constants occurring in two realistic
relativistic point coupling models that have been determined in
Hartree approximation.
The four terms of Eq. (\ref{L1}) are
included, but higher order terms (six- and eight-fermion
point couplings) and derivative terms are included as well.
Here we focus our attention on the four four-fermion point couplings
alone.
 
\noindent The two models are: PC-LA containing 9 coupling constants that appears in
Ref. \cite{Nik92} published in 1992, and PC-F4 containing 11 coupling constants that appears in
Ref. \cite{Bue01} published in 2002. As explained above, the four
four-fermion coupling constants appearing in these tables are taken
as the set \{$\tilde{\alpha}$\} in Eqs. (\ref{L3}) and (\ref{new-alpha}).
\noindent The inverse solution Eq. (\ref{inverse})
then yields
the Hartree-Fock coupling constants \{$\alpha$\}. We show both the
Hartree \{$\tilde{\alpha}$\} and Hartree-Fock
\{$\alpha$\}   coupling constants in Table \ref{tabpc1}.
 
\begin{table}
\caption{Four-Fermion Relativistic Hartree \{$\tilde{\alpha}$\} and Hartree-Fock
\{$\alpha$\} Coupling Constants in Two Realistic Lagrangians (PC-LA and PC-F4)
  [$10^{-4}\;{\rm MeV}^{-2}$].}
\label{tabpc1}
\vspace{12pt}
\begin{tabular}{|l|cccc||cccc|}
\hline
Force & $\tilde{\alpha}_{S}$ & $\tilde{\alpha}_{V}$  & $\tilde{\alpha}_{TS}$ & $\tilde{\alpha}_{TV}$ &
$\alpha_{S}$ &$\alpha_{V}$  & $\alpha_{TS}$ & $\alpha_{TV}$ \\ \hline
PC-LA & -4.508 & 3.427 & 7.421$\times 10^{-3}$ & 3.257 $\times 10^{-1}$ & -5.712 & 4.244 & 2.286 & -2.314 \\
PC-F4 & -3.834 & 2.594 & -5.924 $\times 10^{-2}$ & 3.937$\times 10^{-1}$ & -4.608 & 2.872 & 1.345 & -1.425 \\ \hline
\end{tabular}
\end{table}
 
\noindent Comparing the relativistic Hartree coupling constants of the models PC-LA and PC-F1
with the relativistic Hartree-Fock coupling constants
one observes the following: (a) whereas the exponents in the Hartree
coupling constants range from -7 to -4, those of the Hartree-Fock
are all -4;
(b) the Hartree-Fock isovector-scalar coupling
constant is much larger than its Hartree counterpart,
and has changed sign in PC-F4;
and (c) the Hartree-Fock isovector-vector coupling constant has
changed sign and its absolute magnitude has also increased
in comparison to its Hartree counterpart.
 
\noindent The Hartree-Fock isovector coupling constants have a
larger role than those of Hartree. In fact, the four
Hartree-Fock coupling constants are of the same order of magnitude
and, furthermore, the magnitudes
of the two isovector coupling constants are roughly equal.
But none of this is true in the Hartree case.
The Hartree-Fock coupling constants from both models have the same
signs (not true for the Hartree case).
In addition, it appears from the two models that the sum of the isovector
coupling constants is better determined by the ground state observables
than are the individual values, as was also learned in Ref. \cite{Bue01}.
 
\noindent Finally,
we can ask what are the errors (uncorrelated and correlated)
in the determination of the sets of coupling constants
in the two calculations?
We find that (a) the uncorrelated and
correlated errors in the isovector coupling constants are significantly
diminished in Hartree-Fock approximation (some by roughly two orders of magnitude),
and (b) in this
approximation, all four of the coupling constants are
equally well determined, unlike in the Hartree approximation.
{\em This result is a consequence of the fact that in Hartree-Fock
approximation the four four-fermion coupling constants contribute
in each of the four channels, due to the explicit treatment
of exchange processes.}
 
\noindent Therefore, the
magnitudes of the four four-fermion Hartree-Fock coupling constants,
in the two point-coupling models studied, are well determined and
they are comparable. This immediately brings to mind the question of the
{\em naturalness} of these coupling constants to which we now turn our
attention.
 
\section{The Quest for Naturalness}
The naturalness of the coupling constants relates to the question
as to whether QCD scaling and chiral symmetry apply to finite nuclei.
In 1990, Weinberg \cite{We90} showed that Lagrangians with (broken) chiral symmetry predict
the suppression of N-body forces. He accomplished this by constructing the most general
possible chiral Lagrangian involving pions and low-energy nucleons as an infinite
series of allowed derivative and contact interaction terms and then using QCD energy (mass)
scales and dimensional power counting to categorize the terms of the series. This
led to a systematic suppression of the N-body forces.
 
\noindent We use the scaling procedure of
Manohar and Georgi \cite{MG84} but without pion fields.
Explicit pionic degrees of freedom are absent in RMF Hartree theory,
but can be present in RMF Hartree-Fock theory where Eq. (\ref{IV.3}) then also
contains the pion field and pion mass as in Eq. (1)
of Ref. \cite{FML96}.
The scaled generic Lagrangian term of the (physical) series is, without pions,
\begin{equation}
{\mathcal L} \sim -c_{l n}
\left[ \frac{\overline{\psi}\psi}{f^2_{\pi} \Lambda} \right]^l
\left[ \frac{\partial^{\mu}}{\Lambda} \right]^n
f^2_{\pi} \, \Lambda^2 \,
\label{IV.3}
\end{equation}
 
\noindent where $\psi$ is a nucleon field,
$f_{\pi}$ is the pion decay constant, 92.5 MeV,
  $\Lambda = 770$ MeV is the QCD large-mass scale taken as the $\rho$
meson mass, and ($\partial^\mu$) signifies a derivative.
Dirac matrices and
isospin operators (we use $\vec{t}$ here rather than $\vec{\tau}$)
have been ignored. Chiral symmetry demands \cite{We79}
 
\begin{equation}
\Delta = l + n - 2 \geq 0
\label{IV.2}
\end{equation}
 
\noindent such that the series contains only {\it positive} powers of
(1/$\Lambda$). If the theory is {\it natural} \cite{MG84}, the
Lagrangian should lead to dimensionless coefficients $c_{ln}$ of
order unity.
Our more stringent definition \cite{Bue01} is that a set of QCD-scaled coupling constants
is {\it natural} if their absolute values are distributed about the
value 1 {\it and} the ratio of the absolute maximum value to the absolute
minimum value is {\it less than 10}.
Thus, all information on scales ultimately resides in the
$c_{ln}$. If they are natural, QCD scaling works.
 
\noindent Applying Eq. (\ref{IV.3}) to the
dimensioned relativistic Hartree and Hartree-Fock coupling constants
  of Table \ref{tabpc1}, we
obtain the corresponding sets of QCD-scaled coupling constants listed
in Table \ref{natcc}.
This table shows that,
  whereas the   Hartree coupling constants
are not natural, the   Hartree-Fock coupling constants
are natural. The culprit is the isovector-scalar channel
(corresponding to $\delta$ meson exchange) in the Hartree approximation.
Additional studies with toy Lagrangians consisting {\it only}
of four-fermion interactions, not presented herein, show that
naturalness is recovered in the Hartree-Fock representation from
Hartree solutions where either of the isovector coupling constants
is unusually small.
 The culprit can apparently be {\it either}
of the isovector coupling constants provided their
sum remains approximately constant.
The exchange process, however, dominates in both isovector channels
to the extent that naturalness is recovered in the Hartree-Fock
approximation in either case.
\begin{table}
\centering
\caption{Relativistic Hartree \{$\tilde{\alpha}$\} and Relativistic
Hartree-Fock \{$\alpha$\} Naturalized Coupling Constants \{$c_{ln}$\}
for the Four-Fermion Point Couplings in Two Realistic Lagrangians (PC-LA and PC-F4).}
\vspace{18pt}
\begin{tabular}{|c|c|c|}
\hline
Coup. Const. & $c_{ln}(\rm{PC-LA})$ & $c_{ln}(\rm{PC-F4})$ \\ \hline
$\tilde{\alpha}_{S}$ &     -1.928 &     -1.641    \\
$\tilde{\alpha}_{V}$ &     1.466 &     1.109    \\
$\tilde{\alpha}_{TS}$ &     0.013 &     -0.101    \\
$\tilde{\alpha}_{TV}$ &     0.557 &     0.674    \\
\# natural & 3 & 3 \\
$|\rm{max}|/|\rm{min}|$ & 152. & 16.2 \\ \hline
$\alpha_{S}$ &     -2.443 &     -1.971    \\
$\alpha_{V}$ &     1.815 &     1.229    \\
$\alpha_{TS}$ &     3.912 &     2.301   \\
$\alpha_{TV}$ &     -3.958 &     -2.438   \\
\# natural & 4 & 4 \\
$|\rm{max}|/|\rm{min}|$ & 2.18 &   1.98   \\ \hline
\end{tabular}
\label{natcc}
\end{table}
 
\noindent The explicit inclusion of pion terms in the
Hartree-Fock Lagrangian should not affect these results. This is
because the
generic chiral Lagrangian of Eqs. (11) and (12) has an important
property: refining the model by adding new terms, such as pions,
will change {\it all} of the naturalized coupling constants, but
naturalness will still apply, that is, naturalness is largely
independent of the details, such as adding pions, provided the physics
is introduced via the measured observables in the framework of
these equations.
 
\section{Conclusions}
\noindent We have extracted relativistic Hartree-Fock coupling
constants from the coupling constants for relativistic
Hartree calculations by use of Fierz relations and contact interactions.
Identical observables are calculated with these two approximations,
but the coupling constants and their physical interpretation are different
because, whereas in the Hartree approximation they
{\it implicitly} account for the exchange processes, in the Hartree-Fock
approximation they {\it explicitly} account for the exchange processes.
 
\noindent We
      have learned three things. First, viewing Hartree and Hartree-Fock as
      two different representations we have shown that the smallness of the
      isovector-scalar coupling constant $\alpha_{TS}$ may be an artifact of the
      Hartree representation rather than the signature of a new symmetry
and, therefore, the search for the symmetry is not needed.
Second, while several relativistic Hartree studies have
      concluded that the observables determine a single isovector coupling
      constant (that of the $\rho$--meson), which is natural, we have
      learned   that it is the sum of two
      isovector coupling constants that is preserved in relativistic Hartree and
      that individually these are both natural {\it and} their sum is also preserved in
      relativistic Hartree-Fock. Given that the
       $\delta$--meson  is now playing a role in recent asymmetric nuclear matter
      studies \cite{asynm} it is relevant that
its coupling constant is natural.
Finally, the Hartree-Fock approximation may constitute a physically more
realistic framework for power counting and QCD scaling than the Hartree approximation.
 
\noindent Our conclusions are subject to the assumptions that:
(a)   we are able to neglect tensor
contributions, (b)   our results for
four-fermion contact interactions will not be strongly affected by
Fierz transformations on the remaining higher-order and derivative
terms,
and (c) explicit inclusion of pion interactions will change all of
the coupling constants while retaining their naturalness.
 
\section{Acknowledgements}
This work was supported
 in part by the Bundesministerium f\"ur Bildung und Forschung,
   by Gesellschaft f\"ur Schwerionenforschung,
   and the U. S. Department of Energy.

\end{document}